\documentclass[conference]{IEEEtran}
\usepackage{blindtext, graphicx}
\usepackage{color}

\newcommand{\ie}{i.e.}
\newcommand{\eg}{e.g.}
\newcommand{\etc}{etc.}

\usepackage{cite}

\ifCLASSINFOpdf
\else
\fi
\usepackage[cmex10]{amsmath}
\usepackage{url}
\usepackage{hyperref}
\usepackage{wrapfig}
\usepackage{comment}

\IEEEoverridecommandlockouts

\begin{document}
%
\title{Advertising in the IoT Era:\\Vision and Challenges}


\author{\IEEEauthorblockN{Hidayet Aksu\IEEEauthorrefmark{1}, Leonardo Babun\IEEEauthorrefmark{1}, Mauro Conti\IEEEauthorrefmark{2}, Gabriele Tolomei\IEEEauthorrefmark{2}, and A. Selcuk Uluagac\IEEEauthorrefmark{1}}
\IEEEauthorblockA{\IEEEauthorrefmark{1} Department of Electrical and Computer Engineering\\
Florida International University, Miami, FL, USA\\
Emails:  \href{mailto:haksu@fiu.edu}{{\tt haksu@fiu.edu}}; \href{mailto:lbabu002@fiu.edu}{{\tt lbabu002@fiu.edu}}; \href{mailto:suluagac@fiu.edu}{{\tt suluagac@fiu.edu}}}
\IEEEauthorblockA{\IEEEauthorrefmark{2} Department of Mathematics\\
University of Padua, Italy\\
Emails: \href{mailto:conti@math.unipd.it}{{\tt conti@math.unipd.it}}; \href{mailto:conti@math.unipd.it}{{\tt gtolomei@math.unipd.it}}}
}


%


\maketitle

\begin{abstract}
The \emph{Internet of Things} (IoT) extends the idea of interconnecting computers to a plethora of different devices, collectively referred to as \emph{smart devices}. These are physical items -- \ie, ``\emph{things}'' -- such as wearable devices, home appliances, and vehicles, enriched with computational and networking capabilities.
Due to the huge set of devices involved -- and therefore, its pervasiveness -- IoT is a great platform to leverage for building new applications and services or extending existing ones. In this regard, expanding online advertising into the IoT realm is an under-investigated yet promising research direction, especially considering that traditional Internet advertising market is already worth hundreds of billions of dollars.\\
\indent In this paper, we first propose the architecture of an IoT advertising platform inspired by the well-known business ecosystem, which the traditional Internet advertising is based on. Additionally, 
we discuss 
the key challenges to implement such a platform with a special focus on issues related to architecture, advertisement content delivery, security, and privacy of the users.
\end{abstract}

\begin{IEEEkeywords}
IoT advertising, IoT advertising middleware, IoT ad network, IoT publisher, Internet advertising, Online advertising
\end{IEEEkeywords}

%
\IEEEpeerreviewmaketitle

\section{Introduction}
The Web has gained so much importance in the market economy during the last two decades because of the development of new Internet-based business models. Among those, \emph{online advertising} is one of the most successful and profitable. Generally speaking, online advertising -- also referred to as Internet advertising -- leverages the Internet to deliver promotional contents to end users.
Already in 2011, revenues coming from online advertising in the United States alone surpassed those of cable television, and nearly exceeded those of broadcast television~\cite{IAB2012}.
Plus, worldwide investment in Internet advertising have reached around 200 billion dollars in 2016~\cite{eMarketer2017} and are expected to get to 335 billion by 2020~\cite{eMarketer2016}.\\
\indent Online advertising allows web content creators and service providers -- broadly referred to as \emph{publishers} -- to monetize yet providing their business for free to end users. For example, news websites or search engines can operate without charging users as they get paid by advertisers who compete for buying dedicated slots on those web pages to display ads~\cite{2008-ijeb-jansen, 2009-sigir-broder, 2012-cikm-azimi}.\\
\indent {\color{black}The global spread of mobile devices has also been changing the original target of online advertising \cite{mobileAd, ADVSurvey}. This is indeed moving from showing traditional display advertisements (i.e., \emph{banners}) on desktop computers to the so-called \emph{native} advertisements impressed within app streams of smartphones and tablets~\cite{lalmas2015kdd}.} More generally, Internet advertising business will eventually extend to emerging pervasive and ubiquitous inter-connected \emph{smart devices}, which are collectively known as the \emph{Internet of Things} (IoT).
\\
\indent Enabling computational advertising in the IoT world is an under-investigated research area; nonetheless, it possibly includes many interesting opportunities and challenges. Indeed, IoT advertising would enhance traditional Internet advertising by taking advantage of three key IoT features~\cite{ADVSurvey}: \emph{device diversity}, \emph{high connectivity}, and \emph{scalability}.
IoT device diversity will enable more complex advertising strategies that truly consider \emph{context awareness}. For example, a car driver could receive customized ads from roadside digital advertisement panels based on his habits (\eg, preferred stopping locations, hotels, and restaurants). Furthermore, IoT high connectivity and scalability will allow advertising to be performed in a really dynamic environment as new smart devices are constantly joining or leaving the IoT network. Finally, different from the traditional web browser-based advertising where a limited number of user interactions occur during the day, IoT advertising might count on users interacting with the IoT environment almost 24 hours a day.\\
\indent The rest of this paper is organized as follows: Section~\ref{sec:usecase} motivates the idea of IoT advertising with a use case scenario.
Section~\ref{sec:onlineadvertising} and~\ref{sec:iotoverview} articulate key background concepts.
In Section~\ref{sec:iotadvertising}, we propose our vision of an IoT advertising landscape; in particular, we characterize the main entities involved as well as the interactions between them. Section~\ref{sec:challenges} outlines the key challenges to be addressed for successfully enabling IoT advertising. Finally, we conclude in Section~\ref{sec:conclusions}.

\section{An Example of an IoT Advertising Scenario: In-Car Advertising}
\label{sec:usecase}
\begin{figure}[!htb]
	\centering{\includegraphics[width=0.5\textwidth]{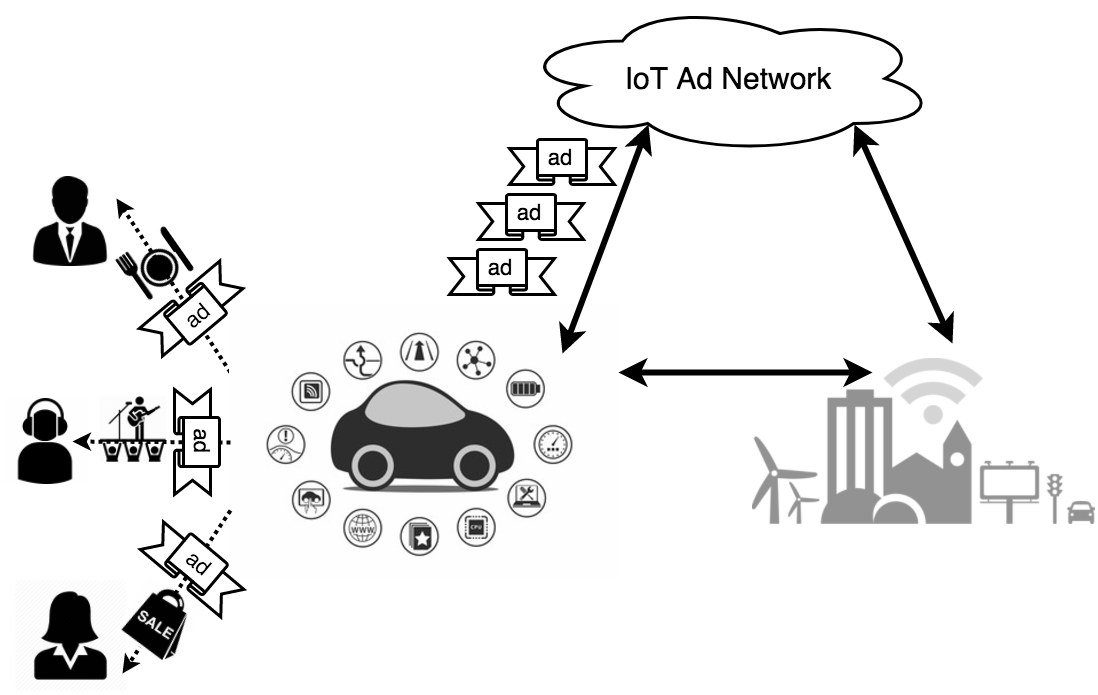}}
	\caption{Targeted ads triggered by the IoT environment (\eg, a smart car traveling close by a smart city) are delivered to end users on IoT devices via an intermediate IoT ad network.}
	\label{fig:usecase}
\end{figure}

Connected smart vehicles are one of the most dominant trends of the IoT industry: automakers are indeed putting a lot of effort to equip their vehicles with an increasing set of computational sensors and devices.\\
\indent With millions of smart vehicles going around -- each one carrying possibly multiple passengers -- automobiles are no longer just mechanical machines used by people to move from point A to point B; rather, they are mobile, interconnected, and complex nodes constituting a dynamic and distributed computing system. This opens up new opportunities for developers who can leverage such an environment to build novel application and services. In particular, smart vehicles -- in fact, passengers traveling on board of those -- may become interesting ``targets'' for advertisers who want to sponsor their businesses.\\
\indent Assume a family of three is traveling in their smart car; their plan is to drive to a seaside destination a few hours away from their home and spend the weekend there. To do so, they rely on the GPS navigation system embedded in their car.
Bob is actually driving the car; he is a forty-five years old medical doctor and he likes Cuban food. Alice -- Bob's wife -- is forty and an architect. She is really passionate about fashion design and shopping.
Sitting in the back of the car, Charlie -- their son -- is a technology-enthusiast teenager who is listening to his favorite indie rock music from his smartphone. Suppose there exists a mechanism for \emph{profiling} passengers traveling on the same smart vehicle, either explicitly or implicitly. In other words, we assume the smart car can keep track of each passenger's profile. {\color{black} Such a profile needs to be built \emph{only} from data which the user agrees to share with the surrounding IoT environment.}
\\
\indent Suppose these travelers are about to cross a city where an iconic summer music festival takes place. Interestingly, an emerging rock band is going to perform on stage the same evening. Festival promoters have already advertised that event through \emph{analog} (\eg, newspapers and small billboards) and \emph{digital} (\eg, the city's website) channels.
However, they would also like to take advantage of an \emph{IoT ad network} to send more targeted and dynamic sponsored messages, namely to reach out to possibly interested people who happen to be around, such as Charlie.\\
\indent Assume Charlie gets an advertisement on the music app installed on his smartphone, and he convinces his parents to stop to attend the concert. Other similar advertising messages might be delivered to Alice and Bob as well. For example, Alice could be suggested to visit the city's shopping mall on her dedicated portion of the car's head-up display. 
Furthermore, the eye-tracking sensors installed in the car could detect that Bob is getting tired, as he has been driving for too long. Therefore, Bob might be prompted with the coordinates of the best local Cuban cafe on the GPS along with a voice message suggesting to have a coffee there.\\
\indent We propose an IoT advertising platform that behaves as an intermediary (\ie, a \emph{broker}) between \emph{advertisers} (the festival promoters), \emph{end-users} (Alice, Bob, and Charlie), and possibly \emph{publishers}, the same way well-known ad networks do in the context of Internet advertising. Note though that in IoT, several entities can play the role of ``publisher'', which is not limited to a single web resource provider, but it may be a composite entity with several IoT devices. As such, the automaker, as well as any other device embedded in the car or dynamically linked to it, may act as publisher. Providing the IoT ad network can gather information from smart vehicles and passengers traveling around a specific geographic area, that information can be further matched against a set of candidate advertisements, which in turn are conveyed to the right target. {\color{black} Note that triggering of ad requests is somewhat transparent to the end user, \ie, we do not conjure any explicit publisher-subscriber mechanism between end users and advertisers. On the other hand, users must have control over their data, which in turn may be used by the IoT ad network for targeting.}\\
\indent Figure~\ref{fig:usecase} depicts the scenario above, where Alice, Bob, and Charlie all receive their targeted advertising messages. The IoT ad network is responsible for choosing the most relevant advertisements and it delivers them through one or more IoT devices that are either embedded in the car (\eg, the head-up display and the GPS) or temporarily joined to the car (\eg, the passengers' smartphones).

{\color{black}We claim that IoT represents a huge opportunity for marketers who may want to leverage the IoT ecosystem to increase their targeted audience. Indeed, although online advertising is already a multibillion-dollar market, we believe one of its limitations is that it is essentially based on the activities users perform on the web. Instead, IoT advertising will overcome this limitation by bringing advertisement messages to users interacting with the IoT environment (which is potentially much larger than the web).}

\section{How Internet Advertising Works Today}
\label{sec:onlineadvertising}
\begin{figure}[!htb]
	\centering{\includegraphics[width=0.46\textwidth]{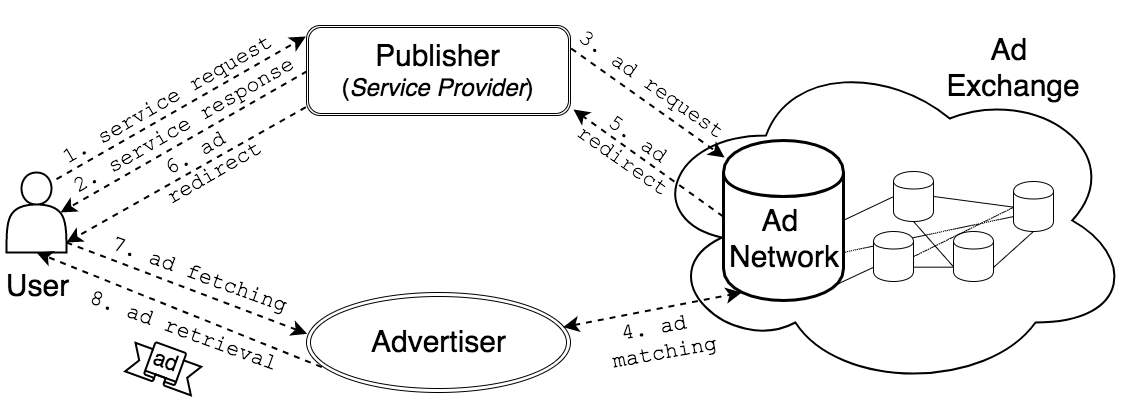}}
	\caption{High-level architecture of traditional online advertising.}
	\label{fig:onlineadvertising}
\end{figure}

The general idea behind Internet advertising is to allow web content \emph{publishers} to monetize by reserving some predefined slots on their web pages to display ads.
On the other hand, \emph{advertisers} compete for taking those slots and are keen on paying publishers in exchange for that.
Actually, publishers often rely on third-party entities -- called \emph{ad networks} -- which free them from running their own ad servers; ad networks decide on behalf of publishers which ads should be placed in which slots, when, and to whom. Furthermore, advertisers partner with several ad networks to optimize their return on investment for their ad campaigns. 
Finally, ad networks charge advertisers for serving their ads according to a specific \emph{ad pricing model}, e.g., \emph{cost per mille impressions} (CPM) or \emph{cost per click} (CPC), and share a fraction of this revenue with the publishers where those ads are impressed \cite{ADVSurvey}.\\
\indent At the heart of online advertising, there is a real-time auction process. This runs within an \emph{ad exchange} to populate an ad slot with an \emph{ad creative}\footnote{\small{An \emph{ad creative} is the actual advertisement message (\eg, text and image) impressed on the slot.}}.
For each ad request, there are multiple competing advertisers bidding for that ad slot.
And, before any ad is served, publishers and advertisers outline a number of ad serving requirements, such as budget, when the ad should be displayed as well as targeting information.
{\color{black}In particular, \emph{targeted advertising} allows to deliver sponsored contents that are more likely tailored to each user's \emph{profile}, which is either explicitly collected (e.g., through the set of user queries submitted to the search engine in the case of sponsored search) or implicitly derived (e.g., from user's browsing history in the case of native advertising)~\cite{yan2009www,li2012dss}.}
The auction process uses all those requirements to match up each ad request with the ``best'' ad creative so as to maximize profit for the publisher.\\
\indent Figure~\ref{fig:onlineadvertising} shows the high-level architecture of current online advertising systems. Although the actual architecture can be more complex than the figure, the main entities involved are: 
the \emph{user} who typically sits behind a web browser or a mobile app; the \emph{publisher} (\ie, a service provider) who exposes some ``service'' to the user (\eg, a web content provider like {\sf \small{cnn.com}} or a web search engine like {\sf \small {Google}} or {\sf \small {Yahoo}}); the \emph{advertiser} who wants to promote its products and possibly attract new customers by leveraging the user base of the publisher; the \emph{ad network} that participates in the \emph{ad exchange} and acts as intermediary between the publisher and the advertiser.

\indent The workflow is as follows: 
\begin{itemize}
\item The user accesses a service exposed by the publisher, \eg, using {\tt HTTP GET} (1).
\item The publisher responses with the ``core'' content/service originally requested (2).
\item The publisher also asks its partner ad network to fetch ads which best match user's profile, and are eventually shown to the user within the same content delivered before (3).
\item The ad network uses profile information during the real-time auction which takes place on the ad exchange to select advertisements that are expected to generate the highest revenue (4).
\item The ad network instructs the publisher on how to tell the user how to fetch the selected ad (5--6).
\item Finally, the user requests (7) and retrieves (8) the actual ad to be displayed.
\end{itemize}

As it turns out from the description above, there is a clear event which activates an ad request, \ie, the user accessing a resource exposed by a web publisher.
Conversely, in the IoT world that triggering event might be less explicit (\ie, the user \emph{interacting} with IoT devices). Nevertheless, in Section~\ref{sec:iotadvertising}, we discuss how the scheme described above can be adapted to the context of future IoT advertising.

\section{IoT Key Features}
\label{sec:iotoverview}
The IoT stack is normally described as a four-layer infrastructure. The first layer defines how the smart physical world (\eg, networked-enabled devices, devices embedded with sensors) interact with the physical world.
The second layer is in charge of providing the necessary connectivity between devices and the Internet. Further, a third layer incorporates data aggregation and other preliminary data processing.
Finally, the fourth layer is in charge of feeding the control centers and providing IoT cloud-based services \cite{IoTArchitecture}. In general, IoT bounds a cooperative relationship among computing systems, devices, and users with these layers.\\ 
\indent \textbf{Connectivity:} A crucial element in IoT is the high connectivity required among devices, servers, and/or service control centers.
Indeed, high-speed connectivity is necessary in order to cope with real-time applications and the level of cooperation expected from IoT devices. Currently, IoT connectivity is guaranteed by traditional network protocols and technologies like WiFi, Bluetooth Smart, and Device-to-Device (D2D) communications. IEEE and the IETF are designing new communications protocols specifically devised for IoT \cite{IOTSurvey}. These protocols (\ie, IEEE 802.15.4e, 6LoWPAN, LoRa) are intended to homogenize the IoT low-energy communication environment among the huge IoT device diversity.\\
\indent \textbf{Resource availability:} This defines the amount of computing resources available to implement IoT services. In general, IoT devices can be categorized into two groups: \textit{resource-rich}, with faster CPUs and higher memory availability and \textit{resource-limited} devices, with limited memory and low-performance CPUs. 
Note that the way IoT devices interact with users (\eg, display availability, user-input enabled devices, \etc) depends on the available resources~\cite{icc17}.\\
\indent \textbf{Power consumption:} The nature of IoT applications imposes several power constraints on the devices. In general, IoT devices are meant to be remotely monitored, autonomous, wearable, and/or with high mobility. These characteristics define the specific power restrictions for every application. 
\\
\indent \textbf{Complexity and Scalability:} Today, IoT devices can be found in several user-oriented (\eg, smart home, wearables devices) and industrial (\eg, smart grid, healthcare IoT) applications.
The different IoT architectures need to be scalable to handle the constant flow of new devices and the always-increasing set of new services and applications.

\section{A Vision for an IoT Advertising Landscape} 
\label{sec:iotadvertising}
\begin{figure}[!htb]
	\centering{\includegraphics[width=0.48\textwidth]{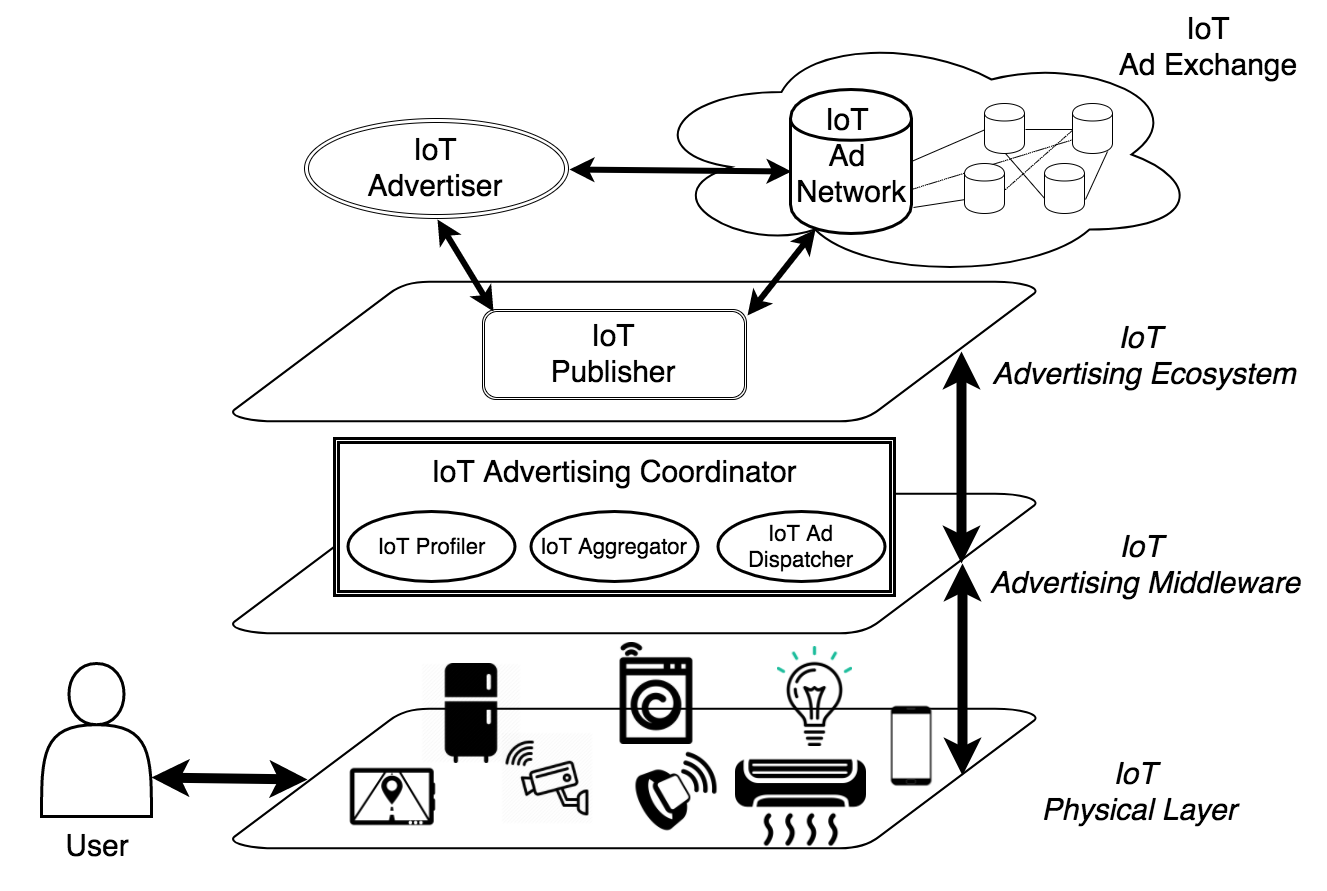}}
	\caption{The proposed IoT advertising model consists of three layers: \emph{IoT Physical Layer}, \emph{IoT Advertising Middleware}, and \emph{IoT Advertising Ecosystem}.}
	\label{fig:iotadvmodel}
\end{figure}

The ultimate aim of IoT is to provide new applications and services by taking advantage of the IoT features discussed above.
Different from the simplistic approach of utilizing traditional legacy sensors combined with decision entities, the high connectivity and intelligence present in IoT along with the possibility of continuous scalability, allow building a wide pool of applications based on users' generated IoT-data. Among those, expanding the traditional Internet advertising marketplace is one of the most promising.\\
\indent To enable the IoT advertisement vision, we introduce our model of an IoT advertising architecture (Figure \ref{fig:iotadvmodel}).
Despite this is clearly inspired by the Internet advertising architecture (Figure~\ref{fig:onlineadvertising}),
IoT advertising has its own peculiarities, and therefore, deserves a dedicated infrastructure to be successful.\\
\indent Our IoT advertising model consists of three layers, each one composed of several entities:
the bottom layer (\emph{IoT Physical Layer}) contains physical IoT devices; the middle layer (\emph{IoT Advertising Middleware}) coincides with the \emph{IoT Advertising Coordinator}, which allows physical IoT devices to interface with the upper layer (\emph{IoT Advertising Ecosystem}), and in particular with the \emph{IoT Publisher}.\\
\indent In the remaining of this section, we discuss the role and characteristics of each entity separately.

\subsection{IoT Advertiser} 
This represents an entity which would like to take advantage of IoT to advertise its own products/services such as the music festival promoters in the use case discussed above. It is expected to interact with other actors of the advertising ecosystem in the same way web advertisers do on traditional Internet advertising. Due to the high diversity of devices involved, the IoT advertiser needs to conceive and design its campaign for heterogeneous targets, \ie, newer ad formats, which are not necessarily visual (\eg, acoustic messages), as opposed to traditional banners displayed on web browsers or mobile apps.
Moreover, targeting criteria may go beyond just user's demographics and/or geolocation; in fact, the contextual environment will play a crucial role in the ad matching phase.

\subsection{IoT Ad Network and IoT Ad Exchange}
The IoT ad network, in combination with the IoT ad exchange will be responsible for matching the most profitable ads with target IoT publishers on behalf of both the publisher and the advertiser. This can be achieved in the same way as traditional ad networks interact with ad exchanges for Internet advertising, \ie, through real-time auctions. Moreover, differently from Internet advertising where those auctions are triggered by the user requesting a resource from a web publisher, in IoT such events can be extremely blurry as the user keeps constantly interacting with her surrounding IoT environment. That means IoT ad networks and ad exchanges may need to operate at an even larger scale and higher rate.

\subsection{IoT Publisher}
The role of IoT publisher is not limited to a web resource provider anymore. An IoT publisher can rather be thought of an ensemble of IoT devices, which collectively cooperate to implement and expose to the user multiple functionalities, as well as to deliver advertisements. For instance, the smart vehicle introduced in our use case is a possible example of an IoT publisher. The smart vehicle is indeed composed of several embedded IoT \emph{atomic} devices (\eg, the GPS, the tire controller, the sound system), each one implementing its own communication standard and exposing a specific functionality through its own user interface. In addition, many other IoT devices can dynamically and temporarily join the smart vehicle (\eg, the smartphones of car passengers).

\subsection{IoT Advertising Coordinator}
The role of IoT advertising coordinator is twofold: On the one hand, it allows bottom-layer IoT devices to expose themselves as a single IoT publisher entity to the upper-layer advertising ecosystem. On the other hand, it is responsible for dispatching and delivering advertisements coming from advertising ecosystem down to physical IoT devices, and in turn, to the end user. To achieve both those capabilities, the IoT advertising coordinator makes use of several sub-components. Among those, we focus on three of them: \emph{(i)} \emph{IoT Aggregator}, \emph{(ii)} \emph{IoT Profiler}, and \emph{(iii)} \emph{IoT Ad Dispatcher}.
Those are responsible for:
\begin{itemize}
\item Unifying different communication standards utilized in a vast variety of IoT devices, so they can all respond to the specific advertisement needs.
\item Providing a cross-platform that will translate IoT-customer interaction into usable data for real-time effective advertisement (\ie, collecting meaningful metadata or ``\emph{profiles}'', which can, in turn, be exploited during ad matching at the layer above).
\item  Managing the actual delivery of advertisements to the target IoT device, and therefore to the user, according to specific supported ad formats.
\end{itemize}
\indent More specifically, the ability of the IoT advertising coordinator to take advantage of IoT devices and user identification via digital fingerprinting will open the door to new advertising strategies. These might consider the following key aspects:
\begin{itemize}
    \item \textit{User profile}: IoT advertising will vary based on the actual recipient (age range, gender, known user behavior) so we can have ads anticipating the user's needs not based on what he/she browses, but based on what he/she is and what he/she does.
    \item \textit{Context awareness}: IoT advertising will adapt to new contexts, that is, the advertisement strategy will also focus on the location, {\color{black}time}, and the type of activity the user is performing (\eg, a regular traveler can receive ads based on the most visited restaurants and hotels {\color{black}during lunchtime}). 
    \item \textit{Services/Features}: IoT advertising ecosystem can make use of an unlimited number of features to know more about the user (\eg, most visited locations, driving mode, behavioral characteristics). That will translate into a new set of services from the IoT advertising landscape (\eg, announcing upcoming events with better price deals, lower car insurance due to the driver record directly derived from the smart car, etc.).
    \item \textit{Security/Privacy}: User security and privacy protection will impact the new IoT advertising model in two different ways. First, the coordinator needs to be transparent to the implementation of traditional (or any new) IoT security mechanisms. Second, these security mechanisms will inevitably limit the amount and type of data that can be extracted from IoT devices and will scarce the quality of the user's digital fingerprint.
    \item \textit{Device capabilities}: The coordinator may have to deal with devices supporting a broader spectrum of advertising formats by themselves (\eg, smartwatches have full display capabilities and adequate computing resources). 
    Conversely, other devices would either accept only custom, resource-friendly ad formats (\eg, acoustic messages sent to smart speakers) or rely on other devices with more capabilities (\eg, the smart lighting system may use the client application running on the smartphone to interact with the user). In this regard, the ad dispatcher will have a crucial role in deciding what specific types of ads to generate/integrate from/to the different devices and how those ads can be delivered to the user.
\end{itemize}

\indent Furthermore, the sensors present in smart devices and interacting with users will play a major role in profiling what the user does (\eg, presence sensor can report when the user leaves the house) and the specific context of such activities (\eg, Saturday night). These constitute key elements for a more effective advertising (\eg, restaurants and nightclubs). Eventually, to be fully effective in a fast-changing and very limited power-consuming IoT world, the amount of data required to characterize users needs to be minimized while coping with the demand imposed by the proposed IoT advertising model. 
In this context, the IoT advertising coordinator will ``translate'' data flow from/to IoT devices into a common language and, more importantly, it will adapt IoT requirements to the well-known Internet advertising model to enable the new IoT advertising ecosystem. Finally, timing and geographical distribution of sensors will influence the effectiveness of the IoT Advertising Coordinator by (1) effectively using user location and IoT device availability to deliver the most appropriated ad (\eg, take advantage of the presence of electronic road signals to show ads to drivers) and (2) timely deliver the right apps (\eg, nearby preferred restaurants at lunchtime).

\section{Challenges of IoT Advertising}
\label{sec:challenges}

In this section, we analyze the possible key challenges of IoT advertising.

\subsection{Architectural Challenges}

From the IoT advertising perspective, the current IoT architecture (see Section~\ref{sec:iotoverview}) has several challenges that need to be addressed. IoT device heterogeneity will add an extra burden to the IoT advertising coordinator. 
The coordinator would need to deal with different memory, CPU, energy, and sensor availability and capabilities, so the right advertising strategy is chosen for every device and user while keeping the required efficiency and reliability of services. Moreover, IoT can be configured in several different network topologies, which require the use of different network metrics to characterize the IoT traffic and to successfully identify devices and users. 

\subsection{Ad Content Delivery Challenges}
Content delivery in IoT advertising involves three different scopes: user profile, user location-activity, and device capabilities. Content delivery challenges will defy the capacity of the IoT devices to cope with the requirements of the proposed IoT advertising scheme in two main aspects: 

\begin{enumerate}
    \item \textit {Quality and quantity of available user data:} Different levels of data obtained from the user will create user-based digital signatures (\ie, user profile) with different quality levels. Also, different permission policies can impact negatively on the quality of users' activity/location tracking processes. 
    \item \textit{Device capabilities:} In cases where IoT device cooperation is not possible, the delivery of the advertisement content to the user will be exclusively defined by the device capacity. For instance, the amount of advertisement content that the user can get from devices with visual capabilities is expected to be higher.
\end{enumerate}

\subsection{Security and Privacy Challenges}
Integrating IoT into the traditional advertising model poses security challenges for customers, advertisers, and publishers.
Some of the security challenges that need to be overcome are the following: 
\begin{itemize}
\item Due to the high diversity of devices and communication protocols in IoT, there exists a perpetual need for monitoring and detecting new vulnerabilities and attacks in a constantly changing environment.
\item Sensitive user data needs to be protected not only from outsiders, but also from malicious corporations that can misuse it.
\item Users are not always aware of security risks and a lot of effort needs to be done on the educational side.
\item Current and new communication protocols incorporate state-of-the-art protection mechanisms, but, in most cases, security is optional and these protocols are insecure in default mode.
\item The high level of interconnection in the IoT opens creates more opportunities for malware and worms to spread over the network.
\item Advertisements should not become intrusive for user privacy nor disrupt the user experience of the surrounding IoT environment.
\end{itemize}
Traditionally, Internet advertising has compromised user privacy by tracking people's browsing habits.
IoT advertising would go further by tracking user behavior based on day-to-day activities. Here, \emph{dataveillance} becomes more valuable considering that IoT user data is much more diverse if compared with regular web browsing data.

\subsection{Fragmentation of IoT} 
Currently, there is not a single inter-operable framework that integrates all IoT devices and services. In fact, despite the efforts to design dedicated protocols for IoT \cite{IOTSurvey}, the current IoT ecosystem offers several options for developers to write smart apps using a variety of different programming architectures (e.g. SmartThings, OpenHAB, and Apple Home Kit). Also, multiple combinations of standards and protocols are possible (\eg, Communications: IPv4/IPv6, RPL, 6LowPAN, Data: MOTT, CoAP, AMPQ, Websocket; Device Management: TR-069, OMA-DM; Transport: Wifi, Bluetooth, LPWAN, NFC; Device Discovery: Physical Web, mDNS, DNS-SD; Device Identification: EPC, uCode, URIs). The proposed IoT advertisement middleware should be able to adapt and convert the current fragmentation of the IoT world into a common language to enable IoT advertising.

\subsection{IoT Data Flow} 
Data flow in IoT highly depends on the programming architecture. There are few cases where smart apps run on specific IoT devices or hubs; however, most of the IoT apps are cloud-based \cite{smartthingspaper}. Smart apps obtain information from the smart devices (sensors) and send data to the cloud to execute the app logic. External web programming tools like IFTTT and Node-RED can also be integrated into the IoT architecture to connect, control, and request information from different devices. The integration of these third-party applications can also represent a challenge to the proposed IoT advertising model. On the other hand, such integration would simplify the overhead caused by the current IoT fragmentation.

\section{Conclusions}
\label{sec:conclusions}
Internet advertising market is worth hundreds of billions of dollars and is one of the fastest growing online businesses. Nevertheless, it is still restricted to web browser-based and, more recently, mobile in-app contexts.\\
\indent The Internet of Things (IoT) will open up a novel, large-scale, pervasive digital advertising landscape; in other words, a new IoT advertising marketplace that takes advantage of a huge collection of smart devices, such as wearables, home appliances, vehicles, and many other connected digital instruments, which end users constantly interact with in their daily lives.\\
\indent In this paper, we introduce the architecture of an IoT advertising platform and its enabling components. We also discuss possible key challenges to implement such a platform with a special focus on issues related to advertisement delivery, security, and privacy of the user.\\
\indent To the best of our knowledge, this is the first work defining the IoT advertising and discussing possible enabling solutions for it.  We expect our work will impact both upcoming researches on this topic, and the development of new products at scale in the industry.


%


\section*{Acknowledgment}
This work is partially supported by the US National Science Foundation (Awards: NSF-CAREER-CNS-1453647, 1663051). Mauro Conti is supported by a Marie Curie Fellowship funded by the European Commission (agreement PCIG11-GA-2012-321980). This work is also
partially supported by the EU TagItSmart! Project (agreement
H2020-ICT30-2015-688061), the EU-India REACH Project (agreement
ICI+/2014/342-896), by the project CNR-MOST/Taiwan 2016-17 ``Verifiable
Data Structure Streaming", the grant n. 2017-166478 (3696) from Cisco
University Research Program Fund and Silicon Valley Community
Foundation, and by the grant "Scalable IoT Management and Key security
aspects in 5G systems" from Intel. The views in this document are of the authors, not of the funding agencies. 

\section*{Additional Note}
\noindent Authors are listed in alphabetical order and each one of them equally contributed to this work.

\ifCLASSOPTIONcaptionsoff
  \newpage
\fi



%

%

\begin{wrapfigure}{l}{0.9in} 
    \includegraphics[width=1in,height=1.25in,clip,keepaspectratio]{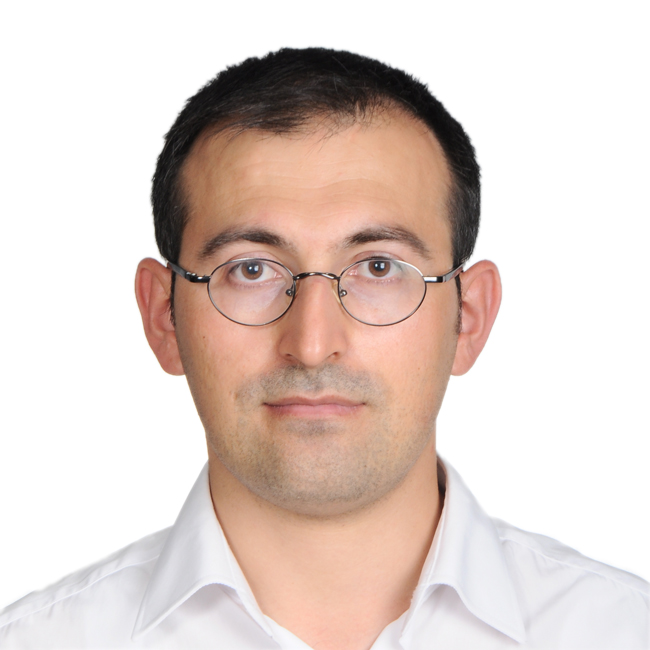}
  \end{wrapfigure} 
\textbf {Hidayet Aksu} received his Ph.D. degree from Bilkent University, in Department of Computer Engineering. He is currently a Postdoctoral Associate in ECE Department of Florida International University. Before that, he conducted research as visiting scholar at IBM T.J.Watson Research Center, USA, for one year.
His research interests include security for cyber-physical systems, internet of things, IoT security, security analytics, social networks, big data analytics, distributed computing, wireless ad hoc and sensor networks, and p2p networks.

\begin{wrapfigure}{l}{0.9in}
    \includegraphics[width=1in,height=1.25in,clip,keepaspectratio]{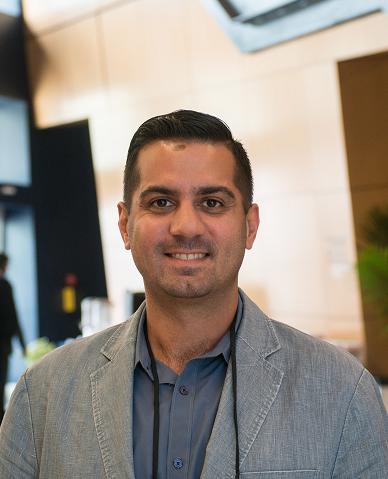}
  \end{wrapfigure} 
\textbf{Leonardo Babun} is currently a PhD student and Research Assistant in the Department of Electrical and Computer Engineering at Florida International University, as a member of the Cyber-Physical Systems Security Lab (CSL).  He previously completed his M.S. in Electrical Engineering from the Department of Electrical and Computer Engineering at Florida International University in 2015. His research interests are focused on Cyber Physical Systems (CPS) and Internet of Things (IoT) security and privacy.

\begin{wrapfigure}{l}{0.9in}
    \includegraphics[width=1in,height=1.25in,clip,keepaspectratio]{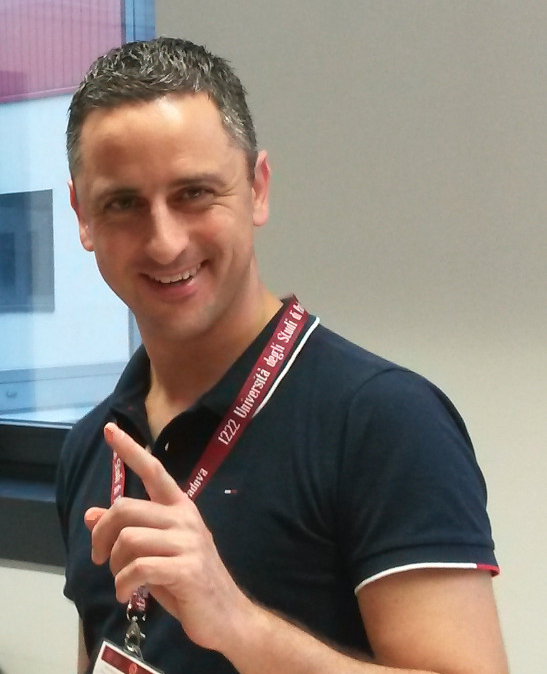}
  \end{wrapfigure} 
\textbf{Mauro Conti} is Associate Professor at the University of Padua, Italy. He
obtained his PhD from Sapienza University of Rome, Italy, in 2009. After
his PhD, he was a Post-Doc Researcher at VU Amsterdam, The Netherlands.
He has been Visiting Researcher at GMU, UCLA, UCI, FIU, and TU
Darmstadt. He has been awarded with a European Marie Curie Fellowship,
and with a German DAAD Fellowship. He is Senior Member of the IEEE.

\begin{wrapfigure}{l}{0.9in}
    \includegraphics[width=1in,height=1.25in,clip,keepaspectratio]{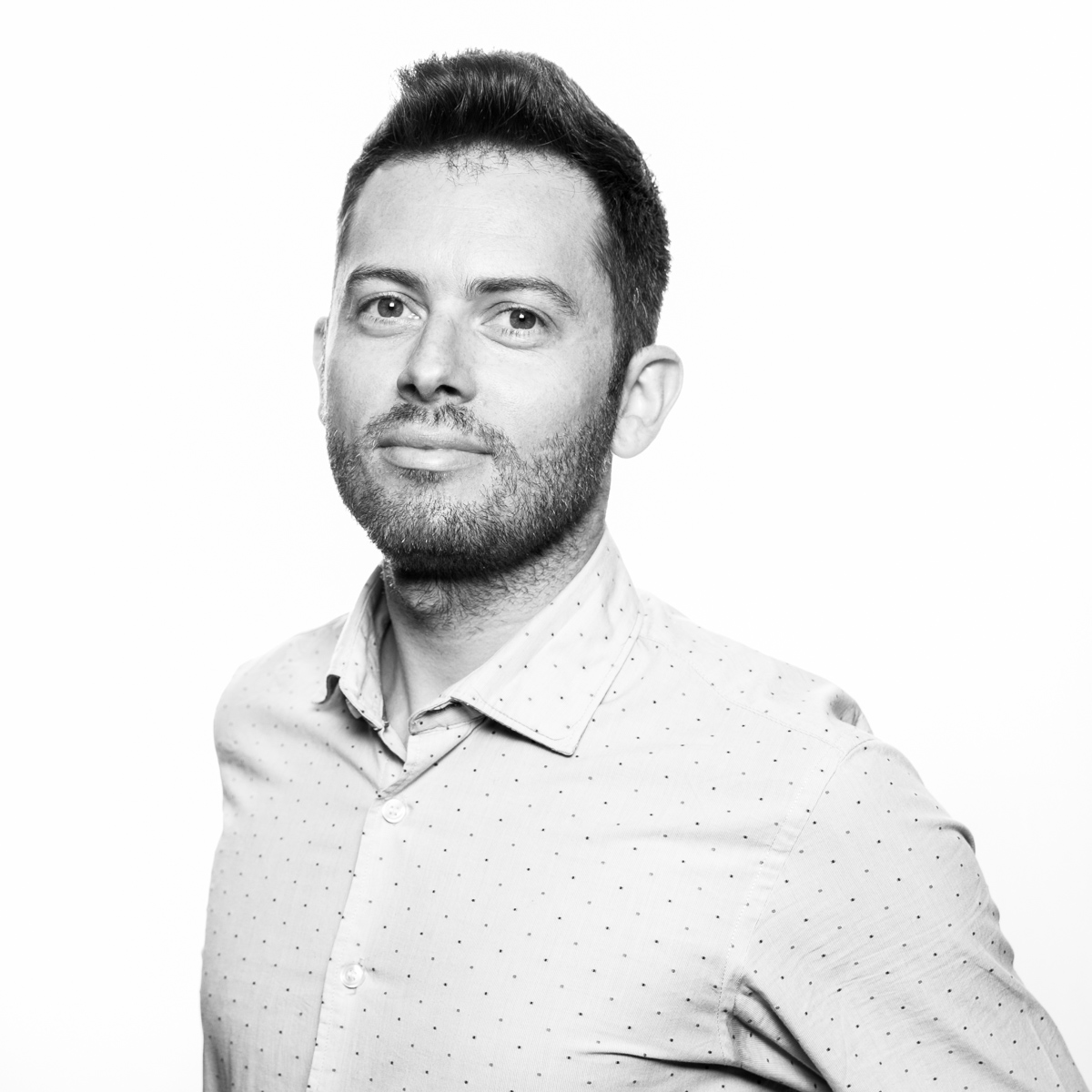}
  \end{wrapfigure} 
\textbf {Gabriele Tolomei} is Assistant Professor at the University of Padua, Italy. Before, he was a Research Scientist at Yahoo Research in London, UK. He received his Ph.D. in Computer Science from Ca' Foscari University of Venice, Italy in 2011. His research interests are: Web Search, Machine Learning, and Computational Advertising. He authored around 30 papers on topmost international journals and conferences, and 4 US patents. He is PC member of many IEEE and ACM conferences.

\begin{wrapfigure}{l}{0.9in}
    \includegraphics[width=1in,height=1.25in,clip,keepaspectratio]{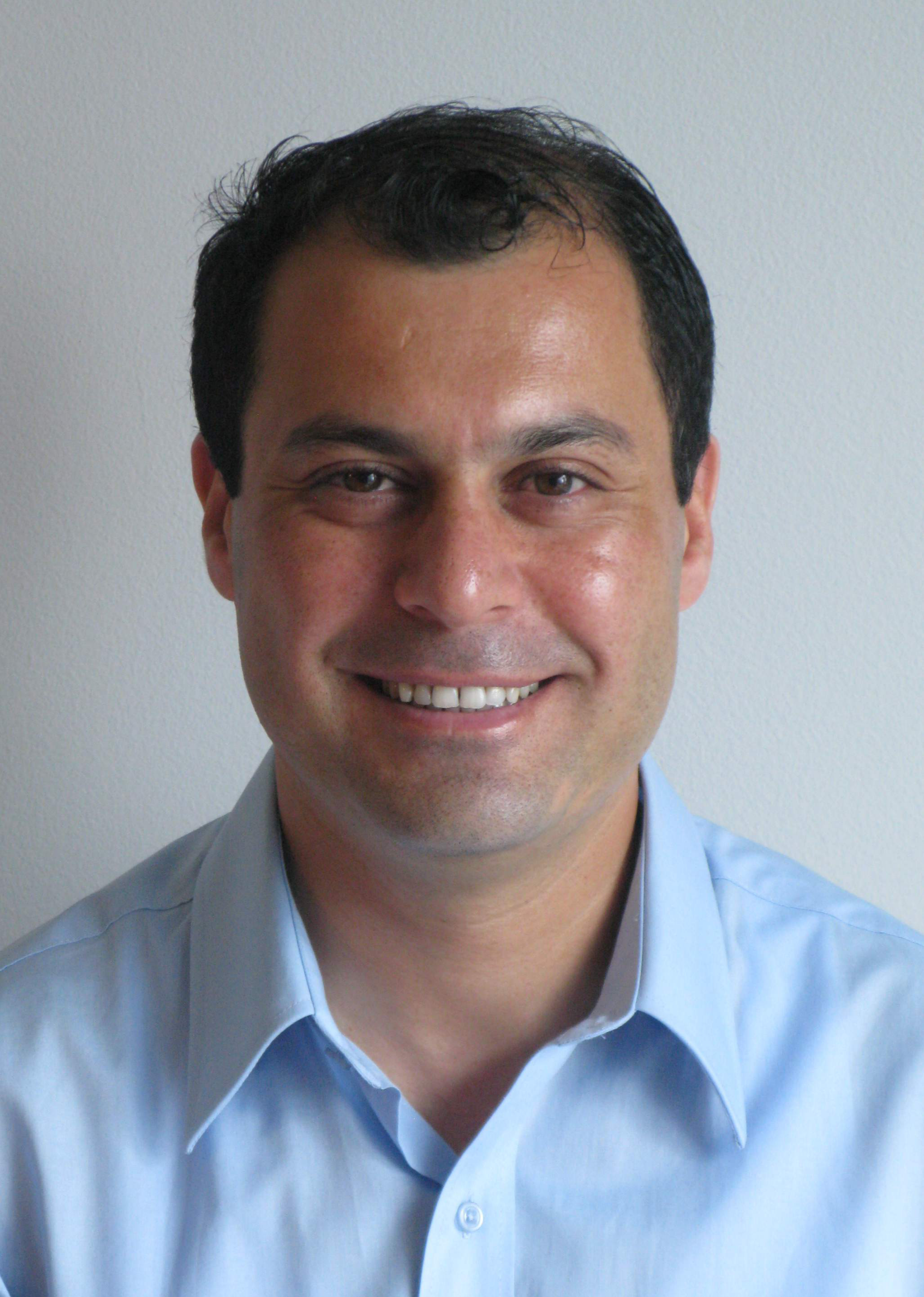}
  \end{wrapfigure} 
\textbf{Selcuk Uluagac} leads the Cyber-Physical Systems Security Lab at Florida International University, focusing on security and privacy of Internet of Things and Cyber-Physical Systems. He has a Ph.D. and M.S. from Georgia Institute of Technology, and M.S. from Carnegie Mellon University. In 2015, he received the US National Science Foundation CAREER award and US Air Force Office of Sponsored Research’s Summer Faculty Fellowship, and in 2016, Summer Faculty Fellowship from University of Padova, Italy.




\end{document}